\documentclass[prl,twocolumn,showpacs,preprintnumbers,amsmath,amssymb,floatfix]{revtex4}

\usepackage{subfigure,graphicx,epsfig,amsmath,amsfonts,amssymb,xcolor,slashed,ulem,multirow}

\usepackage{relsize}
\usepackage{slashed}
\usepackage{color}
\usepackage{ulem}
\usepackage{tabu}

\usepackage{amsmath}
\usepackage{amssymb}
\usepackage{amsthm}
\usepackage{mathrsfs}
\usepackage{graphicx}
\usepackage{epstopdf}
\usepackage{fancyhdr}
\usepackage{array}
\usepackage[all]{xy}
\usepackage{eufrak}
\usepackage{euscript}
\usepackage{enumerate}
\usepackage{slashed}
\usepackage{hyperref}
\usepackage{subfigure} 
\usepackage{epstopdf} 

\usepackage{hyperref}
\hypersetup{pdftex,colorlinks=true,linkcolor=blue,citecolor=blue,menucolor=black,urlcolor=blue,filecolor=blue}

\newcommand{\mev}{\textrm{ MeV}}

\newcommand{\be}{\begin{equation}}
\newcommand{\ee}{\end{equation}}
\newcommand{\ba}{\begin{eqnarray}}
\newcommand{\ea}{\end{eqnarray}}
\newcommand{\nn}{\nonumber}

\begin{document}
\title{Molecular pentaquarks with hidden charm and double strangeness}
\date{\today}

\author{L. Roca}
\email[]{luisroca@um.es}
\affiliation{Departamento de F\'isica, Universidad de Murcia, E-30100 Murcia, Spain}

\author{J. Song}
\email[]{Song-Jing@buaa.edu.cn}
\affiliation{School of Physics, Beihang University, Beijing, 102206, China}
\affiliation{Departamento de F\'{\i}sica Te\'orica and IFIC, Centro Mixto Universidad de
Valencia-CSIC Institutos de Investigaci\'on de Paterna, Aptdo.22085,
46071 Valencia, Spain}

\author{E.~Oset}
\email{oset@ific.uv.es}
\affiliation{Departamento de F\'{\i}sica Te\'orica and IFIC, Centro Mixto Universidad de
Valencia-CSIC Institutos de Investigaci\'on de Paterna, Aptdo.22085,
46071 Valencia, Spain}

\begin{abstract}

We  analyze theoretically the coupled-channel meson-baryon interaction with global flavor $\bar c c s s n$ and $\bar c c s s s$, where mesons are pseudoscalars or vectors and baryons have $J ^P=1/2^+$ or $3/2^+$. 
The aim is to explore whether the nonlinear dynamics inherent in the unitarization process within coupled channels can dynamically generate double- and triple-strange pentaquark-type states ($P_{css}$ and $P_{csss}$  respectively), for which there is no experimental evidence to date.
We evaluate the s-wave scattering matrix by implementing unitarity in coupled channels, using potential kernels obtained from  t-channel vector meson exchange. 
The required $PPV$ and $VVV$ vertices are obtained from Lagrangians derived through appropriate extensions of the local hidden gauge symmetry approach to the charm sector, while capitalizing on the symmetry of the spin and flavor wave function to evaluate the $BBV$ vertex.
We find four different poles in the double strange sector, some of them degenerate in spin. For the triple-strange channel we find the meson-baryon interaction insufficient to generate a bound or resonance state through the unitary coupled-channel dynamics.

\end{abstract}

\maketitle


\section{Introduction}

Over the past decade, pentaquark research has emerged as one of the most active areas in hadronic physics.
A renewed interest was triggered by significant advancements from the
experimental side, namely the report by the LHCb collaboration, since 2015, of various hidden charm pentaquark candidates without strangeness, $P_c(4380)$, $P_c(4312)$, $P_c(4440)$, $P_c(4457)$, $P_c(4337)$
\cite{LHCb:2015yax,LHCb:2019kea,LHCb:2021chn} as well as containing a strange quark,
$P_{cs}(4459)$ \cite{LHCb:2020jpq}, $P_{cs}(4338)$ \cite{LHCb:2022ogu}. 
Numerous theoretical models have competed to offer a theoretical explanation of  their nature (see Refs.~\cite{Chen:2016qju,Lebed:2016hpi,Esposito:2016noz,Guo:2017jvc,Ali:2017jda,Liu:2019zoy,Chen:2022asf} for some reviews).
Particularly satisfactory have been the meson-baryon molecular interpretations on its nature, with some of these states predicted \cite{Wu:2010jy,Wu:2010vk,Wang:2011rga,Yang:2011wz,Wu:2012md,Xiao:2013yca,Li:2014gra,Chen:2015loa,Karliner:2015ina} before their discovery by LHCb.
These interpretations rely on implementing unitarity in coupled channels starting from potentials grounded in t-channel vector meson exchange in most cases.

A natural step forward would be to inquire about the potential existence of hidden-charm pentaquarks with strangeness $S=-2$ or $S=-3$, namely $P_{css}$ and $P_{csss}$, with quark content $\bar c c s s n$ and $\bar c c s s s$ respectively. Such hypothetical states are yet to be substantiated by experimental findings.
However, for the $S=-2$ channel, no theoretical evidence of resonant or bound states was found in \cite{Wu:2010vk,Xiao:2013yca} using as coupled channels $D_s\Xi'_c$, $D_s\Xi_c$ and $\bar D \Omega_c$. The reason for the null result was  explained in Ref.~\cite{Marse-Valera:2022khy}, where it was argued that while the $\bar D \Omega_c$ to $\bar D \Omega_c$ potential is small by itself, it was crucial to obtain enough attraction via the nonlinear interaction with other non-diagonal terms of the potential, neglected in \cite{Wu:2010vk,Xiao:2013yca}. Notably, 
one pole for the $PB$ interaction, with mass about $4493\mev$ and width $74\mev$, and another one for $VB$ scattering, with mass around $4633\mev$ and width $80\mev$, were found in \cite{Marse-Valera:2022khy}.
In Ref.~\cite{Wang:2020bjt}, also within a molecular picture albeit a different formalism to \cite{Marse-Valera:2022khy}, poles were also found but using somewhat large regulator cutoffs.
Other theoretical approaches regarding $P_{css}$ pentaquarks are based on sum rules \cite{Azizi:2021pbh} and quark models \cite{Anisovich:2015zqa,Ortega:2022uyu}.
Even more scarce is the theoretical study of possible triple-strange hidden-charm pentaquarks \cite{Wang:2020bjt,Wang:2021hql,Azizi:2022qll}.

In the present work we  evaluate the meson-baryon interaction, where the mesons are either pseudoscalars or vectors, and baryons have $J ^P=1/2^+$ or $3/2^+$, with a total flavour $\bar c c s s n$ and $\bar c c s s s$.
We implement unitarity in coupled channels building upon a kernel from potentials derived from t-channels vector meson exchange.
In addition to the consideration of the $3/2^+$ baryons and the calculation  of the $S=-3$ channels, the main difference with respect to the work in \cite{Marse-Valera:2022khy} is the simpler way we evaluate the $BBV$ vertex, by considering directly the flavor and spin wave functions with their proper symmetrization, as done in \cite{Debastiani:2017ewu,Wang:2022aga}, in contrast to Ref.~\cite{Marse-Valera:2022khy} where $SU(4)$ symmetry is invoked.

The unitarized scattering amplitudes considered in the present work reveal four distinct poles across the various channels examined within the double-strange sector which can be associated to pentaquark states of $\bar c c s s n$ flavor, (or $P_{css}$). Conversely, we find that the interaction in triple-strange channels lacks the strength necessary to yield poles.

\section{Formalism}

The formalism for the evaluation of the meson-baryon interaction in s-wave relies upon implementing unitarity in coupled channels, following the techniques of the chiral unitary approach \cite{Kaiser:1995eg,Oset:1997it,Oller:1998zr,Oller:2000fj,Dobado:1996ps,Oller:1998hw}. 
We consider in the present work the following meson-baryon sets of channels:

\begin{align}
&PB[\bar c c ssn](\tfrac{1}{2}^-) :  \eta_c \Xi(4298), \bar D_s\Xi'_c(4546),
  \bar D \Omega_c(4560)  \nn\\
&  VB[\bar c c ssn](\tfrac{1}{2}^-,\tfrac{3}{2}^-) :  J/\Psi \Xi(4411), \bar D_s^*\Xi'_c(4690),\nn \\
&\qquad\qquad\qquad\qquad\quad\
  \bar D^* \Omega_c(4702)\nn \\
&  PB^* [\bar c c ssn](\tfrac{3}{2}^-):  \eta_c \Xi^*(4515), \bar D_s\Xi^*_c(4613),
  \bar D \Omega^*_c(4630) \nn\\
&  VB^* [\bar c c ssn](\tfrac{1}{2}^-,\tfrac{3}{2}^-,\tfrac{5}{2}^-): J/\Psi \Xi^*(4628), \bar D_s^*\Xi^*_c(4757), \nn \\
&\qquad\qquad\qquad\qquad\qquad\quad \ \  \bar D^* \Omega^*_c(4773) \nn \\
&    PB^* [\bar c c sss](\tfrac{3}{2}^-): \eta_c \Omega(4656), \bar D_s\Omega^*_c(4734) \nn\\
&     VB^* [\bar c c sss](\tfrac{1}{2}^-,\tfrac{3}{2}^-,\tfrac{5}{2}^-)
   :  J/\Psi\Omega(4769), \bar D^*_s\Omega^*_c(4878) \nn
\end{align}

\noindent where $P$ stands for pseudoscalar meson, $V$ for vector meson, $B$ for baryon with $J^P=\frac{1}{2}^+$ and $B^*$ for a $\frac{3}{2}^+$ baryon. In square brackets the flavor content of the channel is represented (where $n$ stands for either $u$ or $d$), followed by the possible $J^P$ values of the meson-baryon channel (which are degenerate in our model). The number in parenthesis following a particular channel represents its threshold energy in MeV. The channels with $\frac{3}{2}^+$ baryons, $B^*$, were not considered in Refs.~\cite{Wu:2010vk,Marse-Valera:2022khy}, whereas they were also taken into account in Ref.~\cite{Wang:2020bjt}.
We do not consider the mixing of  different sets of channels, such as $PB$ and $VB$, as the $PB$ to $VB$ transition should proceed through pseudoscalar exchange, which is subdominant to the vector exchange \cite{Dias:2021upl}.
On the other hand, the $\bar D_s\Xi_c$ channel  is neglected in the $PB$ set since the $\Xi_c$ flavor and spin wave functions have opposite symmetry to the rest of baryons (see for instance Table IV in Ref.~\cite{Wang:2022aga}) and our $BBV$ vertex is spin independent in the limit of small three-momentum. Note that in the model of Ref.~\cite{Marse-Valera:2022khy} this channel is included in their formalism but the coupling to the other channels is very suppressed. In our formalism, it could have been evaluated as an independent channel but it turns out to be repulsive and hence has no chance to generate a bound state or resonance. An analogous reasoning applies to the 
$\bar D_s^*\Xi_c$ channel for $VB$ interaction.
Similarly, in the $S=-3$ sector, the $PB$ channel $\bar D_s\Omega_c$ does not mix with the $PB^*$ ones and its self-transition is positive, resulting in no attraction.

\begin{figure}[h]
\centering
\includegraphics[width=.45\linewidth]{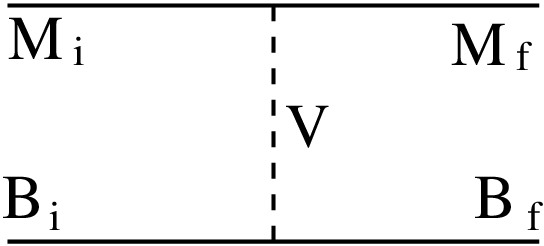}
\caption{Tree level potentials through vector meson exchange.}
\label{fig:figdiag1}
\end{figure}

The $MB$ tree level interaction within each channel set proceeds through the exchange of a vector meson between the meson and the baryon, as depicted in Fig.~\ref{fig:figdiag1}, for which we need the vertices $VPP$, $VVV$, $BBV$ and $B^*B^*V$. 
In order to account for these vertices
we use a formalism similar  to \cite{Xiao:2019gjd,Feijoo:2022rxf,Wang:2022aga} based on heavy
quark spin symmetry together with the local hidden
gauge approach (LHG) \cite{Bando:1984ej,Bando:1987br,Birse:1996hd,Meissner:1987ge,Nagahiro:2008cv} extrapolated to the charm sector.
The LHG formalism is one of the most successful realizations of chiral symmetry involving vector mesons. In this framework, vector meson fields are regarded as gauge bosons of a hidden local symmetry, undergoing inhomogeneous transformations, and represent the most natural means to incorporate vector meson dominance.
Extending the LHG framework to encompass the charm \cite{Molina:2009ct,Molina:2010tx} and beauty quark sectors \cite{Ozpineci:2013qza,Dias:2014pva,Molina:2010tx} has proven highly effective in addressing meson-meson and meson-baryon interactions involving both hidden and open charm and beauty mesons and baryons. Additionally, studies such as those in Refs.~\cite{Xiao:2013yca,Liang:2014eba} have shown that LHG preserves the heavy quark spin symmetry, a feature of QCD in which the interaction of heavy quarks remains independent of their spin.

Within the LHG formalism, the $VPP$ and $VVV $ Lagrangians read

\begin{eqnarray}\label{eq:VPP}
  {\cal{L}}_{VPP} &=& -ig \,\langle [P, \partial_\mu P] V^\mu\rangle \,,\nn \\
  {\cal{L}}_{VVV} &=& ig \,\langle (V^\mu \partial_\nu V_\mu-\partial_\nu V^\mu V_\mu) V^\nu\rangle \,,\label{eq:VVV}
\end{eqnarray}
 with $g =\frac{M_V}{2\,f}$ for which we take $M_V=800 \mev$, $f=93 \mev$, and $P$ and $V$ are $q\bar q$ matrices (considering $u$, $d$, $s$ and $c$ quarks)  expressed in terms of mesons:
 

 \begin{equation}
\label{eq:matP_charm} 
  P = \begin{pmatrix}
         0 & 0 & 0 & \bar{D}^0 \\
          0 &0 & 0 & D^- \\
          0 & 0 & 0 & D_s^- \\
          D^0  & D^+ & D_s^+ & \eta_c
       \end{pmatrix},
\end{equation}

\begin{equation}
  \label{eq:matV_charm}
   V = \begin{pmatrix}
             \frac{1}{\sqrt{2}}\rho^0 + \frac{1}{\sqrt{2}} \omega & \rho^+ & K^{* +} & \bar{D}^{* 0} \\
             \rho^- & -\frac{1}{\sqrt{2}}\rho^0 + \frac{1}{\sqrt{2}} \omega  & K^{* 0} & \bar{D}^{* -} \\
             K^{* -} & \bar{K}^{* 0}  & \phi & D_s^{* -} \\
             D^{* 0} & D^{* +} & D_s^{* +} & J/\psi
        \end{pmatrix}.
\end{equation}
where we have shown in the $P$ matrix only the pseudoscalar mesons needed in the present work.

For the $BBV$ interaction we adopt an analogous formalism to \cite{Debastiani:2017ewu,Wang:2022aga} (see those references for further details), and this is one of the main differences from the approaches in  Ref.~\cite{Wu:2010vk,Marse-Valera:2022khy}.
 Instead of employing an effective Lagrangian, the approach involves expressing the 
wave function of the baryons in terms of quarks, implementing the corresponding flavor symmetry of each baryon and hence considering the spin flavour to ensure overall wave function symmetry, (see table  IV in \cite{Wang:2022aga}). For these states, rather than using $SU(4)$ symmetry, we single out the heavy quarks and impose the symmetry on the light quarks, following the approach of Refs.~\cite{Capstick:1986ter,Roberts:2007ni}
This formalism allows also for a very similar evaluation of the $B^*B^*V$ vertex, which were not considered in \cite{Wu:2010vk,Marse-Valera:2022khy} since it would have required the use of much more involved effective Lagrangians.
 The vertex is then given by \cite{Debastiani:2017ewu,Wang:2022aga}.

\be
\widetilde{\cal L}_{VBB}=g\, q \bar q\, \gamma^\mu\epsilon_\mu,
\ee
where $q\bar{q}$ is the flavor wave function of the vector meson, and we keep only the time component of $\gamma^\mu\epsilon_\mu\sim \gamma^0\epsilon^0\sim \epsilon^0$, valid for slow baryons.

The tree level transition potentials for the $MB$ mechanisms depicted in Fig.~\ref{fig:figdiag1} take the following form (see Appendix in Ref.~\cite{Debastiani:2017ewu}  for an explicit example of a detailed evaluation for a different case):

\begin{eqnarray}
  \label{eq:def_Vij}
     V_{ij}=g^2 C_{ij}(p^0+p'^{0}) \, ,
\end{eqnarray}
where $p^0(p'^0)$ are the on-shell center of mass energy of the initial(final) meson, and the coefficients $C_{ij}=C_{ji}$ are given in Tables~\ref{Tab:CijPBVPu} to \ref{Tab:CijPBVPsss}. 
Note the dependence of the coefficients on the inverse of the mass squared of the exchanged vector meson in the t-channel, for which we use the mass of the actual meson exchanged in each particular channel. This  is also a refinement with respect to \cite{Marse-Valera:2022khy} where the mass of the light vectors were assumed identical, those with one charm were twice the mass of the light ones, and the mass of the $J/\Psi$ was three times that of the light ones.

In Table~\ref{Tab:CijPBVPu} we show the $C_{ij}$ coefficients for the $PB$ channels with $S=-2$.
\begin{table}[h!]
  \begin{center}
    \begin{tabular}{c|cccc}
        \hline\\ [-0.30cm]
         & $\eta_c\Xi$ &  $\bar{D}_s\Xi_c^\prime$ & $\bar{D}\Omega_c$\\
        \hline \\ [-0.2cm]
        $\eta_c\Xi$ & $0$ & $\frac{1}{\sqrt{6}m^2_{D^*_s}}$ & $-\frac{1}{\sqrt{3} m^2_{D^*}}$ \\ 
        $\bar{D}_s\Xi_c^\prime$ & &$\frac{1}{m^2_\phi}-\frac{1}{m^2_{J/\Psi}}$ & $\frac{\sqrt{2}}{m^2_{K^*}}$ \\
        $\bar{D}\Omega_c$ & & & $-\frac{1}{m^2_{J/\Psi}}$ \\
    \end{tabular}
  \end{center}
\caption{$C_{ij}$ coefficients of the $PB$ interaction in the $\bar c c s s n$ sector.}    
\label{Tab:CijPBVPu}
\end{table}
For the $VB$ channel the table is the same, except for substituting the pseudoscalar by the corresponding vector meson.
The coefficients for the channels involved in the double-strange $PB^*$  interaction are given in Table~\ref{Tab:CijPBstarVPu}, which are also the same for $VB^*$ if we replace again the name of the pseudoscalar by the vector meson.
\begin{table}[h!]
  \begin{center}
    \begin{tabular}{c|cccc}
        \hline\\ [-0.30cm]
         & $\eta_c\Xi^*$ &  $\bar D_s\Xi^*_c$ & $\bar D \Omega^*_c$\\
        \hline \\ [-0.2cm]
        $\eta_c\Xi^*$ & $0$ & $\sqrt{\frac{2}{3}}\frac{1}{m^2_{D^*_s}}$ & $\frac{1}{\sqrt{3} m^2_{D^*}}$ \\ 
        $\bar D_s\Xi^*_c$ & &$\frac{1}{m^2_\phi}-\frac{1}{m^2_{J/\Psi}}$ & $\frac{\sqrt{2}}{m^2_{K^*}}$ \\
        $\bar D \Omega^*_c$ & & & $-\frac{1}{m^2_{J/\Psi}}$ \\
    \end{tabular}
  \end{center}
\caption{$C_{ij}$ coefficients of the $PB^*$ interaction in the $\bar c c s s n$ sector.}    
\label{Tab:CijPBstarVPu}
\end{table}

For the $S=-3$ sector the coefficients $C_{ij}$ are given in Table~\ref{Tab:CijPBVPsss} for the $PB^*$ interaction, and  analogous for $VB^*$, changing the pseudoscalar by the corresponding vector meson.
\begin{table}[h!]
  \begin{center}
    \begin{tabular}{c|cc}
        \hline\\ [-0.30cm]
         & $\eta_c \Omega$ &  $\bar D_s\Omega^*_c$\\
        \hline \\ [-0.2cm]
        $\eta_c \Omega$ & $0$ & $\frac{1}{m^2_{D^*_s}}$  \\ 
        $\bar D_s\Omega^*_c$ & &$\frac{2}{m^2_\phi}-\frac{1}{m^2_{J/\Psi}}$ \\
     \end{tabular}
  \end{center}
\caption{$C_{ij}$ coefficients of the $PB^*$ interaction in the $\bar c c s s s$ sector.}    
\label{Tab:CijPBVPsss}
\end{table}

It is worth noting that the coefficients in the first row of Tables~\ref{Tab:CijPBVPu} to \ref{Tab:CijPBVPsss} are suppressed by the squared mass of a heavy vector meson. However, since the first channel is the only one open at the energy where the pole of the generated states appears, it is the only source of the width of the generated resonance. Despite this, it has minimal influence on the mass of the generated state.

It is interesting to comment on the analogies between the matrix
elements obtained in Ref.~\cite{Marse-Valera:2022khy} and those in
 Table~\ref{Tab:CijPBVPu} (see Table I
of \cite{Marse-Valera:2022khy}). The diagonal terms coincide (beware of a global sign difference).
The large nondiagonal term for $\bar{D}_s\Xi_c^\prime$ to $\bar{D}\Omega_c$ transition, $\sqrt{2}$,
also coincides (there is a misprint in the sign of this term in Table I
of~\cite{Marse-Valera:2022khy}) but the transitions from $\bar{D}_s\Xi_c^\prime$ and $\bar{D}\Omega_c$ to $\eta_c\Xi$
are a factor $-1/\sqrt{3}$ smaller in our case. This difference is
important, because the only decay channel for the states that we obtain
is $\eta_c\Xi$, which means that we expect to obtain much smaller widths
than in Ref.~\cite{Marse-Valera:2022khy}.  These features were already observed in the study of
the $\Omega_c$ states in \cite{Montana:2017kjw} and \cite{Debastiani:2017ewu}. The  large terms in the
matrix elements (the 1 and the $\sqrt{2}$) come from the exchange of
light vectors. In this case the heavy quarks are spectators and one is
projecting $SU(4)$ to $SU(3)$ and the matrix elements obtained are the same
in both schemes. Yet, when it comes to the exchange of heavy vectors,
our approach and the use of $SU(4)$ give different results, and that was
the case in \cite{Montana:2017kjw} and  \cite{Debastiani:2017ewu}, as well as in the present study. These differences do
not matter much in the evaluation of the masses of the states, but  they
are relevant in the width which goes trough terms involving the exchange
of heavy vectors in the present case.

Within the framework  of the coupled channels unitary approach, exact unitarity can be incorporated into the meson-baryon interaction using as kernels the tree level potentials of Eq.~\eqref{eq:def_Vij}. To this aim, we use the Bethe-Salpeter equation, which is analogous to the $N/D$ \cite{Oller:1998zr,Oller:2000fj} or IAM \cite{Dobado:1996ps,Oller:1998hw} formalisms:

\begin{equation} T=(1-VG)^{-1}V \ ,
\label{eq:BS}
\end{equation}
where we have factored out a global $\Vec{\epsilon}_i\cdot\Vec{\epsilon}_j$ term for the channels involving a vector meson in the external states and after neglecting factors of order  $\frac{q^2}{M_V^2}$ in internal vector propagators.
In Eq.~\eqref{eq:BS}, $G$ represents a diagonal matrix containing the meson-baryon loop functions:
\begin{eqnarray}
        G_l(\sqrt{s}) & =&  \int_{q< \Lambda} \frac{d^3q}{(2\pi)^3}\frac{1}{2\omega_l({q})}\frac{M_l}{E_l({q})} \nonumber\\
                &\cdot&  \frac{1}{\sqrt{s}-\omega_l({ q}) - E_l({\vec q})+i\epsilon}\,,
          \label{eq_Gl}
\end{eqnarray}
where $q=|\vec q|$, $\omega_l(q)=\sqrt{m^2_l+q^2}$ and $E_l=\sqrt{M^2_l+q^2}$
with $m_l$ and $M_l$ representing the mass of the meson and baryon in the loop, respectively. To control the logarithmic divergence of the loop function, regularization is necessary. We opt for the cutoff method, bounding the three-momentum integral to a value $\Lambda=600$~MeV, as used in Ref.~\cite{Feijoo:2022rxf} in the generation of $P_{cs}$ pentaquarks with strangeness -1 and similar to the value of 
650~MeV used in Ref.~\cite{Debastiani:2017ewu} in the generation of molecular $\Omega_c$ states. 
The value of this cutoff is the only free parameter of our model but it has a natural size linked to the dynamical scale integrated out in the model, such as the mass of the lighter vector mesons exchanged in the t-channel diagram. It is, hence, the main source of uncertainty of the results. We estimate the uncertainty from this source by considering also  the value  $\Lambda=800\mev$ used in Ref.~\cite{Marse-Valera:2022khy}, since it would be a measure of the sensitivity of our results on the regularization parameter.
The loop functions can also be regularized by means of dimensional regularization but, 
although both regularization techniques typically yield equivalent outcomes, several works \cite{Wu:2010rv,Xiao:2013jla,Ozpineci:2013qza,Feijoo:2022rxf} have shown that in the heavy flavor domain the cutoff approach proves more suitable. Indeed, and in addition to the naturalness of the regularization scale discussed above, the cutoff method respects heavy quark symmetry since the three-momentum cutoff value remains independent of the heavy flavor \cite{Lu:2014ina,Altenbuchinger:2013vwa}. 

Should the meson-baryon interactions investigated in this study give rise to pentaquark resonances, they would manifest as poles in the $T_{ij}$ scattering amplitudes on the second Riemann sheet. If a pole, at $\sqrt{s_R}$, is not very far from the real axis, it can be associated to the mass, $M$, and width, $\Gamma$, of a resonance as $\sqrt{s_R}=M-i \Gamma/2$. 
On the other hand the couplings of the generated resonance to a given channel, $i$, can be evaluated from the residue at the pole of the amplitude, since close to the pole the dominant term of the Laurent expansion of the amplitude is

\begin{equation}
T_{ij}=\frac{g_i g_j}{\sqrt{s}-\sqrt{s_R}}.
\end{equation}

In addition we can evaluate the compositeness, $X_i$, of a generated resonance for a given channel, $i$, which represents the weight of a molecular component in the wave function of the state, as \cite{Gamermann:2009uq,Hyodo:2011qc,Aceti:2014ala}
\begin{eqnarray}
X_i=-g_i^2 \left. \frac{\partial G_i}{\partial \sqrt{s}}\right |_{\sqrt{s}}.
\label{eq:Xi}
\end{eqnarray}
Higher compositeness would suggest a more molecular nature for the resonance, while lower values may indicate a more compact structure.


\section{Results}

\begin{table*}[htbp]\centering\small%
\caption {Results for the mass, $M$, width, $\Gamma$,  couplings, $g_i$, and the compositeness $|X_i|$ for the different channels. 
The upper number in the numerical cells represent the value obtained with the  cutoff $\Lambda=600\mev$ and the lower one using $\Lambda=800\mev$.
(Masses and widths in MeV). }
\begin{tabular}{|ccc|c|c|c|c|c|c|}%
\hline
     & flavor &           $I(J^P)$                     & $M$                                                  & $\Gamma$   &     &  $g_i$  & $|g_i|$ & $|X_i|$                   \\ \hline
$PB$ & $\bar c c s s n$ &  $\frac{1}{2}(\frac{1}{2}^-)$ & \begin{tabular}{@{}c@{}} 4535 \\ 4479 \end{tabular} &   \begin{tabular}{@{}c@{}} 9 \\ 12 \end{tabular}         & $\eta_c \Xi$ & \begin{tabular}{@{}c@{}}  0.39+i0.01\\ 0.64-i0.00 \end{tabular} &\begin{tabular}{@{}c@{}}  0.39\\0.64 \end{tabular} &\begin{tabular}{@{}c@{}}  0.01\\0.03 \end{tabular} \\ \cline{6-9} 
&&&&& $\bar D_s\Xi'_c$ & \begin{tabular}{@{}c@{}} 
 -1.47-i0.16\\ -2.56-i0.07 \end{tabular} &\begin{tabular}{@{}c@{}}  1.48\\2.56 \end{tabular} &\begin{tabular}{@{}c@{}}  0.44\\0.38 \end{tabular} \\ \cline{6-9}
&&&&& 
  $\bar D \Omega_c$ & \begin{tabular}{@{}c@{}}  2.12+i0.20\\ 3.42+i0.09 \end{tabular} &\begin{tabular}{@{}c@{}}  2.13\\3.42 \end{tabular} &\begin{tabular}{@{}c@{}}  0.53\\0.59 \end{tabular} \\ \hline
  
 $VB$ & $\bar c c s s n$ &  $\frac{1}{2}(\frac{1}{2}^-,\frac{3}{2}^-)$ & \begin{tabular}{@{}c@{}} 4675 \\ 4617 \end{tabular} &   \begin{tabular}{@{}c@{}} 10 \\ 12 \end{tabular}         
     & $J/\Psi \Xi$      & \begin{tabular}{@{}c@{}}  0.41+i0.01\\ 0.66-i0.00 \end{tabular} &\begin{tabular}{@{}c@{}}  0.41\\0.66 \end{tabular} &\begin{tabular}{@{}c@{}}  0.01\\0.04 \end{tabular} \\ \cline{6-9} 
&&&&& $\bar D_s^*\Xi'_c$ & \begin{tabular}{@{}c@{}}  -1.59-i0.17\\ -2.70-i0.08 \end{tabular} &\begin{tabular}{@{}c@{}}  1.60\\2.70 \end{tabular} &\begin{tabular}{@{}c@{}}  0.42\\0.38 \end{tabular} \\ \cline{6-9}
&&&&& 
  $\bar D^* \Omega_c$   & \begin{tabular}{@{}c@{}}  2.25+i0.21\\ 3.58+i0.11 \end{tabular} &\begin{tabular}{@{}c@{}}  2.26\\3.58 \end{tabular} &\begin{tabular}{@{}c@{}}  0.56\\0.60 \end{tabular} \\ 
  \hline 
  
  $PB^*$ & $\bar c c s s n$ &  $\frac{1}{2}(\frac{3}{2}^-)$ & \begin{tabular}{@{}c@{}} 4602 \\ 4548 \end{tabular} &   \begin{tabular}{@{}c@{}} 0 \\ 0 \end{tabular}         
     & $\eta_c \Xi^*$      & \begin{tabular}{@{}c@{}}  0.02+i0.00\\ 0.01+i0.00 \end{tabular} &\begin{tabular}{@{}c@{}}  0.02\\0.01 \end{tabular} &\begin{tabular}{@{}c@{}}  0.04\\0.00 \end{tabular} \\ \cline{6-9} 
&&&&& $\bar D_s\Xi^*_c$ & \begin{tabular}{@{}c@{}}  -1.42-i0.00\\ -2.48-i0.00 \end{tabular} &\begin{tabular}{@{}c@{}}  1.42\\2.48 \end{tabular} &\begin{tabular}{@{}c@{}}  0.45\\0.38 \end{tabular} \\ \cline{6-9}
&&&&& 
  $\bar D \Omega^*_c$   & \begin{tabular}{@{}c@{}}  2.13-i0.00\\ 3.37-i0.00 \end{tabular} &\begin{tabular}{@{}c@{}}  2.13\\3.37 \end{tabular} &\begin{tabular}{@{}c@{}}  0.49\\0.57 \end{tabular} \\ 
  \hline 
 
   $VB^*$ & $\bar c c s s n$ &  $\frac{1}{2}(\frac{1}{2}^-,\frac{3}{2}^-,\frac{5}{2}^-)$ & \begin{tabular}{@{}c@{}} 4743 \\ 4686 \end{tabular} &   \begin{tabular}{@{}c@{}} 0 \\ 0 \end{tabular}         
     & $J/\Psi \Xi^*$      & \begin{tabular}{@{}c@{}}  0.02+i0.00\\ 0.00+i0.00 \end{tabular} &\begin{tabular}{@{}c@{}}  0.02\\0.00 \end{tabular} &\begin{tabular}{@{}c@{}}  0.02\\0.00 \end{tabular} \\ \cline{6-9} 
&&&&& $\bar D_s^*\Xi^*_c$ & \begin{tabular}{@{}c@{}}  -1.56-i0.00\\ -2.63-i0.00 \end{tabular} &\begin{tabular}{@{}c@{}}  1.56\\2.63 \end{tabular} &\begin{tabular}{@{}c@{}}  0.43\\0.38 \end{tabular} \\ \cline{6-9}
&&&&& 
  $\bar D^* \Omega^*_c$   & \begin{tabular}{@{}c@{}}  2.27-i0.00\\ 3.54-i0.00 \end{tabular} &\begin{tabular}{@{}c@{}}  2.27\\3.54 \end{tabular} &\begin{tabular}{@{}c@{}}  0.52\\0.58 \end{tabular} \\ 
  \hline

   \hline

\end{tabular}
\label{tab:results1}
\end{table*}

The results of our analysis for the mass, $M$, width, $\Gamma$, couplings, $g_i$, and the compositeness $|X_i|$ for different channels are presented in Table~\ref{tab:results1}. The upper values in the numerical cells correspond to the calculations with a cutoff $\Lambda=600\mev$, while the lower values represent the results for $\Lambda=800\mev$.
 As discussed in the formalism section, the difference between these two numbers reflects the sensitivity of the results to the regularization parameter and can be taken as an estimate of the uncertainty of our calculation. However, by analogy to the work done with the $P_{cs}$ states \cite{Feijoo:2022rxf}, we favor the result with $\Lambda=600\mev$.
For the strangeness $S=-3$ channels we do not find poles, and therefore, there are no results to be shown in the table for them.

Going into the details of the results, for the $PB$ channel with flavor content $\bar c c s s n$ (that is, $J^P=\frac{1}{2}^-$, isospin $I=1/2$,  $S=-2$) we predict a resonance with a mass in the range $M=4479-4535\mev$ and width $\Gamma=9-12\mev$, considering the uncertainty from the cutoff dependence. It is worth comparing this result with the value obtained in Ref.~\cite{Marse-Valera:2022khy}, $M=4493\mev$ and $\Gamma=74\mev$. While the mass aligns with our results within the uncertainty, our result for the width is significantly smaller. This discrepancy mainly stems from the fact that there is roughly a $-1/\sqrt{3}$ factor difference in the $\eta_c\Xi$ couplings to the other channels  compared to \cite{Marse-Valera:2022khy}  (first row in Table~\ref{Tab:CijPBVPu}), from where roughly a factor 3 reduction in the width would result. Indeed, as discussed in the formalism section, this channel is responsible for the finite imaginary part of the amplitude. This $-1/\sqrt{3}$ factor comes from the different formalism used, as was discussed above.
A similar comparison and discussion can be made for the $VB$ channel. In our analysis we find, for the dynamically generated state, a mass range of $M=4617-4675\mev$
and $\Gamma=10-12\mev$, degenerate in spin $(\frac{1}{2}^-,\frac{3}{2}^-)$, and in Ref.~\cite{Marse-Valera:2022khy} a value of $M=4633\mev$ and $\Gamma=80\mev$ was obtained.

We can also see from the values of the coupling strengths to the different channels, that the state generated in the $PB$ channel couples dominantly to $\bar D \Omega_c$. This channel, along with $\bar D_s\Xi'_c$, has a significant weight (compositeness) in the wave function of the generated resonance. Consequently, the resulting state can predominantly be characterized as a molecular state of the $\bar D \Omega_c$, $\bar D_s\Xi'_c$ components.
An analogous conclusion can be drawn for the $VB$ channel.

We also show in Table~\ref{tab:results1} the resulting values of $|X_i|$. Actually, $X_i$ is complex but the imaginary part is negligible in the present case. This still allow us to interpret $|X_i|$ as probabilities,
consistently with the interpretation of the complex $X_i$ in  the case of open channels discussed in detail in Ref.~\cite{Aceti:2014ala}.
Note that the sum of probabilities for the closed channels in Table~\ref{tab:results1} is of the order of 0.94-1.00 in all cases, indicating a clean molecular structure.

For the channels involving $J^P=\frac{3}{2}^+$ baryons, not considered in \cite{Marse-Valera:2022khy}, namely $PB^*$ and $VB^*$, our analysis predicts two exceedingly narrow resonances with flavor  $\bar c c s s n$ and $I(J^P)=\frac{1}{2}(\frac{3}{2}^-)$ with mass $M=4548-4602\mev$ for the $PB^*$ state, and $M=4686-4743\mev$ for the $VB^*$ with  $I(J^P)=\frac{1}{2}(\frac{1}{2}^-,\frac{3}{2}^-,\frac{5}{2}^-)$. In these cases, also the strongest coupling and the dominant molecular weight correspond to the heaviest channel, $\bar D \Omega^*_c$ and $\bar D^* \Omega^*_c$ respectively.
Note that the coupling strengths of these states to  the lowest mass channels, $\eta_c \Xi^*$  or $J/\Psi \Xi^*$, are minimal, resulting in a negligibly small width for the generated states.
Another interesting observation is that the poles are consistently located at a similar distance below the threshold of the dominant channel, amounting to approximately $25\mev$ for $\Lambda=600$ and $85\mev$ for $\Lambda=800$.

Regarding the hidden-charm $S=-3$ channel, no poles are found of the unitarized scattering amplitudes, indicating a lack of resonant or bound state behavior in those channels. This can be understood within our model by taking into account the specific coefficients in  Table~\ref{Tab:CijPBVPsss}, all of which are positive. For these channels the $V$ matrix takes the form
\begin{equation}
\label{eq:matV} 
  V = \begin{pmatrix}
         0 & V_{12}  \\
          V_{12} & V_{22}
       \end{pmatrix}.
\end{equation}

While it is true that all elements in the $V$ matrix are positive, this does not necessarily rule out the possibility of a global attraction in the whole couple-channel dynamics. Attraction could still arise if the non-diagonal term of the potential were sufficiently large. However, in the present analysis, this is not the case.
 Indeed, in order to have a pole of the $T$ matrix in  Eq.~\eqref{eq:BS}, we need that
\be
det(1-VG)=0\ \Rightarrow \ 1- V_{22}G_2 - V_{12}^2 G_1 G_2=0.
\label{eq:det0}
\ee
Note that $G_2<0$ and $V_{22}>0$ in this case, which implies $-V_{22}G_2>0$ and cannot contribute to cancel the term '1'. However,  there is still a possibility of finding a solution to Eq.~\eqref{eq:det0} if  $V_{12}^2 G_1 G_2$, which is $>0$, is big enough to cancel the remaining $1- V_{22}G_2$ term. However, we have checked that in the present case $V_{12}^2 G_1 G_2$ is about two orders of magnitude smaller than $V_{22}G_2$.
Equivalently, $V_{22} + V_{12}^2G_1$ can be interpreted as an effective potential for the channel 2, which can only be negative (attractive) if $V_{12}$ is sufficiently large, (bear in mind that the real parts of the loop functions are negative).
This negative conclusion concerning $S=-3$ states contrasts with the one of Ref.~\cite{Wang:2021hql}, where some states are obtained in particular cases using relatively large regulator cutoffs.

The previous argument underscores the intricate nonlinear dynamics inherent in the unitarization process. Another illustration of the non-trivial behavior of the unitarization was previously emphasized in Ref.~\cite{Marse-Valera:2022khy} concerning the $S=-2$ channels. Indeed, for these channels, attraction primarily originates from the strong non-diagonal term $C_{23}$ in Table~\ref{Tab:CijPBVPu} or \ref{Tab:CijPBstarVPu}, without which the $C_{33}$ term would lack sufficient strength to generate a pole by itself.
The importance of coupled channels to generate states in the $S=-2$ sector was also emphasized in Ref.~\cite{Wang:2020bjt}.
 If we consider the analogue of Eq.~\eqref{eq:det0}, neglecting the weak $\eta_c \Xi$ channel, we encounter the pole condition 
$1- (V_{33} + V_{23}^2 G_2) G_3=0$. In this case, $V_{33}$ is attractive but too small to cancel out the term '1' by itself, requiring the contribution of $V_{23}^2 G_2$ to provide enough attraction to get a solution of Eq.~\eqref{eq:det0} and consequently a pole in the $T$ matrix.
It is also worth noting that in another study aimed at generating these states \cite{Dong:2021juy}, only the exchange of $\phi$ mesons was considered, with no inclusion of coupled-channels. As a result, no poles were found in \cite{Dong:2021juy} as the single channel ($\bar D_s \Xi'_c$) considered is repulsive, overlooking the significance of the smaller $\bar D \Omega_c$ channel and  the  crucial cross-interaction between these channeles, as discussed above.

\section{Conclusions}

Motivated by the experimental discovery of a pentaquark with hidden charm and single strangeness, $P_{cs}$, we have studied theoretically the possible existence of states with hidden charm and double or triple strangeness, $P_{css}$ and $P_{csss}$ respectively. To this end, we evaluate the meson-baryon scattering amplitudes, with mesons being pseudoscalars or vectors and the baryon having $J ^P=1/2^+$ or $3/2^+$, with total flavor $\bar c c ss n$ and $\bar c c ss s$. Our approach implements the techniques of the chiral unitary approach to resum the coupled channels multiple final state interaction,  inherent to the unitarization procedure, starting from kernel potentials based on t-channel vector meson exchange. The required $VPP$ and $VVV$ vertices are obtained from suitable Lagrangians provided by the local hidden gauge symmetry approach, properly extended to the charm sector in a way validated  in numerous previous studies.  For the $BBV$ and $B^*B^*V$ vertices we use a simplified model
that capitalizes on  the symmetry of the spin and flavor wave function of the intervening hadrons, which is notably simpler, albeit accurate, than previous far more complicated Lagrangians used in the literature.

By searching for poles in unphysical Riemann sheets of the unitarized scattering amplitudes, we find four distinct poles for the different channels considered in the double-strange sector, which can be associated to four different pentaquark-like states
of $\bar c c s s  n$ flavor, $P_{css}$. However, for the triple-strange channels, we discuss that the interaction is not strong enough to generate poles. 

It is important to emphasize that the emergence of the poles is solely a result of the intricate nonlinear dynamics inherent in the unitarization in coupled channels via the Bethe-Salpeter equation. 
These poles manifest themselves without the necessity of incorporating them as explicit degrees of freedom, relying solely on the input of tree-level potentials. 
The only freedom of the model, and thus the main source of uncertainty, is the value of the three-momentum cutoff which acts as the regulator of the logarithmically divergent meson-baryon loop function inherent to the unitarization formalism.

Despite the inherent uncertainty, the conclusion drawn from the current study remains robust and well-founded: these pentaquark-type  double-strange hidden-charm  states are practically compelled to exist and it ought to encourage experimental efforts aimed at their discovery.

\section*{ACKNOWLEDGEMENTS}

This
work is partly supported by the Spanish Ministerio
de Economia y Competitividad (MINECO) and European FEDER funds under
Contracts No. FIS2017-84038-
C2-1-P B, PID2020-112777GB-I00, and by Generalitat
Valenciana under contract PROMETEO/2020/023. This
project has received funding from the European Union
Horizon 2020 research and innovation programme under
the program H2020-INFRAIA-2018-1, grant agreement
No. 824093 of the STRONG-2020 project.
This work of J. S. is partly supported by the National Natural Science
Foundation of China under Grants No. 12247108 and the China Postdoctoral
Science Foundation under Grant No. 2022M720360 and  No. 2022M720359.


\end{document}